# On the Warming of the Southern Hemisphere since 1955 and Recent Slowing-Down: Role of Sea Ice, Sun Spot and El Nino Variability

*Alfred Laubereau and Hristo Iglev*

Physik-Department E11, Technische Universität München, James-Franck-Strasse, D-85748 Garching, Germany

ABSTRACT. The importance of the sea ice retreat in the polar regions for the global warming and the role of ice-albedo feedback was recognized by various authors [1,2]. Similar to a recent study of the phenomenon in the Arctic [3] we present a semi-quantitative estimate of the mechanism for the Southern Hemisphere (SH). Using a simple model, we estimate the contribution of ice-albedo feedback to the mean temperature increase in the SH to be 0.5 ± 0.1 K in the years 1955 to 2015, while from the simultaneous growth of the greenhouse gases (GHG) we derive a direct warming of only 0.2 ± 0.05 K in the same period. These numbers are in nice accordance with the reported mean temperature rise of 0.75 ± 0.1 K of the SH in 2015 since 1955 (and relative to 1880). Our data also confirm previously noticed correlations between the annual fluctuations of solar intensity and El Nino observations on the one hand and the annual variability of the SH surface temperature on the other hand. Our calculations indicate a slowing down of the temperature increase during the past few years that is likely to persist. Assuming a continuation of the present trends for the southern sea ice and GHG concentration we predict the further temperature rise to decrease by 33 % in 2015 to 2025 as compared to the previous decade.

INTRODUCTION:

The present climate change has different impacts on the two polar regions [1]. While the Arctic is warming at a greater rate than elsewhere [2,4-6], the Antarctic experiences predominantly cooling [7]. Sea ice plays an important role in this development. In contrast to the continuous northern ice retreat [8], the ice surrounding Antarctica has grown since 1979 when satellite-based measurements began [9-11]. This trend is surprising as compared with the general warming of the globe and the prediction of many climate models. Most of the increase has occurred in the western Ross Sea whereas the sea ice cover in the Bellinghausen Sea is shrinking [12,13]. Several regional and global processes are discussed in the literature that influence sea ice growth and melt, e.g. tropical Pacific and Atlantic teleconnections, variability in the winds and ocean currents around Antarctica including stratospheric ozone depletion [14]. The deep ocean can have an outsized effect on the Southern Ocean because it is a region of significant upwelling leading to a delayed response of the southern sea ice to the general warming [7]. The ice-albedo feedback is widely recognized as an important feature in this context [15]. A recent attempt to quantify the impact of sea ice on the terrestrial climate concluded that its influence on the global heat budget has similar magnitude as the emissions of greenhouse gases [16]. The present paper suggests an even larger effect and underlines the important role of sea ice for the terrestrial climate.

In this paper the reported ice changes in the Antarctica are taken as empirical facts and used to compute the corresponding temperature variation. A historical record exists on the sea ice extent prior to the satellite period so that the data for the sea ice area may be extended using proxies and data assimilation. Since 1979 the ice areas are known from satellite observations including the phase shift of annual melting and freezing relative to the seasonal solar input [10]. The data used in the present study are depicted in Fig. 1 (black circles) supplemented by linear fits (straight solid lines) [17,18]. The figure shows that in 1979 to 2015 the ice maxima increase from $15.14 \times 10^{12}$ to $15.86 \times 10^{12}$ m², while in the Antarctic summer the minima rise from $1.65 \times 10^{12}$ to $2.06 \times 10^{12}$ m² (fitted values). The values for the sea ice extent show a similar behavior in that period as indicated by the red triangles in the figure [18]. For earlier years only data on the sea ice extent are available (see Fig. 1) [19,20]. Assuming that the ratio between ice area and ice extent is approximately constant in 1955 to 1979 we use the extent data to determine the ice area values for this time interval, arriving at values of $19.6 \times 10^{12}$ m² and $5.0 \times 10^{12}$ m² for maximum and minimum area in 1955 (see solid black lines in Fig. 1). The sea ice retreat from 1955 to 1979 is noteworthy. The numbers represent a decrease of maximum (minimum) ice area of approximately 23 % (67 %) from 1955 to 1979. Until 2015 the losses of sea ice area are estimated to be 19 % and 59 %, respectively for Antarctic winter and summer. Of course, the accuracy of these numbers is limited. The values are consistent with Ref. [7] that sea ice coverage might have been up to 25 % in the 1940s to 1960s.

Taking into account the smaller backscattering of sea water compared to snow-covered ice for incident sun light (decrease by a factor of 0.5) and averaging over the daily and seasonal variations of the solar input, the net input of solar radiation in the Antarctic is estimated to grow by $1.85 \times 10^{14}$ W in 1955 to 2015. Relative to the total solar input of the SH in 1955 ($8.73 \times 10^{16}$ W) we arrive at a decrease of the global albedo (0.300 in 1955) of 0.7 % (see Appendix 1).

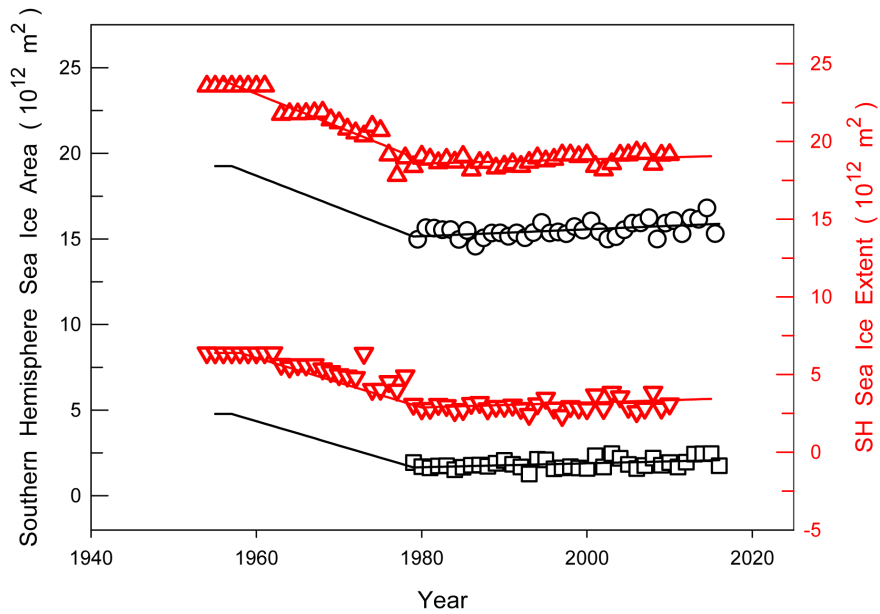

**Figure 1.** Retreat of the Antarctic sea ice in the years 1955 to 1979 and subsequent minor increase until 2015: the annual maxima and minima values of sea ice area are plotted (experimental black circles and fitted solid black line, left hand ordinate scale). Auxiliary data for the annual maximum and minimum values of sea ice extent are included (reported red triangles and fitted red lines, right hand ordinate scale). The extent data are used to estimate the sea ice area in 1955 -1979 assuming that the ratio of sea ice area to sea ice extent is constant in 1955 to 1979; data taken from Ref. [17-20]; see text.

We have developed a simple 1-dimensional model for the causal effect of albedo changes for the mean SH surface temperature including also the effect of greenhouse gases (GHG). The same model was previously used [3] and is based on the spectroscopic properties of the GHG and several empirical facts, e.g. solar input and albedo factor. In a simplified picture, the atmosphere is represented by four layers with mean temperatures $T_j$ ($j = 1 - 4$). The approach is an extension of the well-known 2-layer model for the greenhouse effect [21-23]. The surface is represented as a black radiator with intensity $\sigma \cdot T_0^4$ (Stefan-Boltzmann constant $\sigma$) while the atmospheric parts are treated as selective thermal emitters according to the spectral properties. The temperatures $T_{surf}$, $T_j$ are self-consistently determined from thermal quasi-equilibrium between the solar input and the thermal emissions of the surface and atmospheric layers without a fitting parameter. For details see Appendix 2.

We consider the 1-dimensional treatment quite satisfactory in the sense of a Taylor expansion of the (unknown) function $T_{surf}$ depending on a variety of local and geographical quantities and the concentrations of greenhouse gases. Second and higher order terms of that expansion are neglected. Thus retaining only the first order terms, the problem is linearized and averaging over the manifold of geographical parameters and GHG properties can be interchanged leading to a 1-D treatment. GHG concentrations are taken from Refs. [24] and [25].

RESULTS AND DISCUSSION:

Because of the lack of sufficiently precise measurements of the global albedo factor the albedo changes have to be estimated from the variations of the sea ice area (see Appendix 1). The resulting surface temperature change $\Delta T_{surf}$ as predicted by our model is plotted in Fig. 2 (theoretical curves) together with the reported time evolution of the mean surface temperature change of the Southern Hemisphere (open gray circles). The latter information is taken from the literature [26]. Only part of that experimental data (starting in 1880) is shown in the Fig. from 1930 onwards, arbitrarily up-shifted by 0.15 K so that $\Delta T_{surf,exp} \sim 0.0$ K in 1880. There is a clear trend in the reported annual temperature variations superimposed by experimental inaccuracies that will be discussed below. For the years 1955 to 2015 the well-known dramatic rise of $\Delta T_{surf,meas}$ by $0.75 \pm 0.1$ K is indicated (smaller than in the NH).

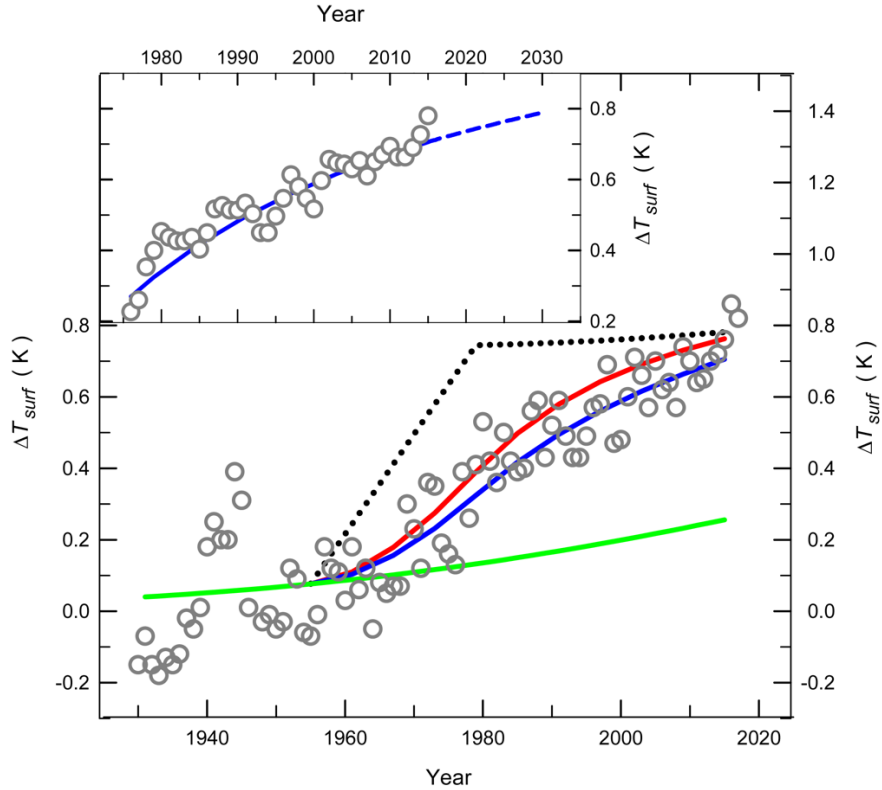

**Figure 2.** Calculated change of the SH mean surface temperature, $\Delta T_{surf}$ and measured data $\Delta T_{surf,exp}$ (open gray circles); the latter are taken from Ref. [26] but arbitrarily upshifted by 0.15 K so that the mean temperature change in 1880 is approximately set to zero (not shown in the Fig.). The reported temperature rise by $\approx 0.75$ K until 2015 is noteworthy. Green curve: computed temperature change resulting directly from the spectroscopic properties of the green house gases including water for constant albedo = 0.300 and constant sea ice area. The combined effect of the direct GHG contribution and of the Antarctic sea ice melting in 1955 - 2015 is shown for three situations: (i) SH warming for instantaneous equilibration of the ice contribution (dotted black line); (ii) delayed equilibration of the temperature rise via ice loss with time constant $\tau = 20$ yr (red curve) and (iii) with $\tau = 30$ yr (blue line).
Inset: Temperature change $\Delta T_{surf}$ in 1975 to 2030. To reduce experimental scatter 3-year averages of $\Delta T_{surf,exp}$ are plotted (open grey circles, r.h.s ordinate scale). The calculated curve for equilibration with $\tau = 30$ yr is also shown (blue curve). Extrapolation for 2015 to 2030 yields the dotted blue line with $\Delta T_{surf} = 0.06$ K in 2015 to 2025; see text.

The computed results for the temperature rise by the greenhouse gases (including water) as derived from our model for constant albedo (i.e. constant sea ice area) are shown in Fig. 2 by the green curve. The spectral forcing via the changing spectral properties of the GHG with concentration is considered; for details see [23]. A minor increase of 0.08 K is evaluated for 1955 relative to 1880 (see solid green line). A further rise of 0.18 K is computed for 1955 to 2015 because of the concentration increase of the GHG (broken green curve). Comparison with the experimental data (open dark gray circles) readily shows that the increasing far infrared absorption of the GHG with growing abundance is not the major cause for the warming of the SH surface. At this point it is interesting to compare our data with the reported spectral forcing of the GHG of 1.82 W/m$^2$ for the years 1750 - 2015 [27]. For a $CO_2$ concentration of 275 ppmV in 1750 we calculate a temperature rise of 0.29 K by the GHG corresponding to a spectral forcing of 1.57 ± 0.19 W/m$^2$ in nice accordance with the published number.

The dominant role of the Antarctic sea ice for the temperature rise of the SH surface, as unraveled by our investigation, is indicated by the additional data in Fig. 2. As mentioned above, the feedback of the albedo factor on the surface temperature is included by computations. To this end the lowering of the albedo factor by the melting of Antarctic sea ice has to be determined. We estimate the change to be -0.00212 (-0.71 % of 0.300) for 1955 to 2015 if instantaneous response of the mean SH value to the increase of solar input in the Antarctic is assumed. The corresponding temperature rise of the SH surface in that case is evaluated to be 0.78 K in 1955-2015 (dotted black line in Fig. 2), including the GHG contribution. The calculated curve differs from the measured data (compare experimental points in the Fig). For an explanation of the deviation we recall that a prompt equilibration of the global temperature to changes in the Antarctic may not be expected. In fact, the well-known permanent temperature difference between equator and pole area indicates a rather slow thermal equilibration in the southern hemisphere.

As a consequence, we extend our model and incorporate a delayed response of the mean albedo in the SH to changes in the Antarctica by the help of an equilibration time $\tau$ (see Appendix 1). The results for time constants of respectively $\tau = 20$ years (red curve) and 30 years (blue line) are presented in Fig. 2. The time required for equilibration in the SH delays the corresponding global warming with a minor reduction of the temperature rise until 2015 to 0.71 K for $\tau = 30$ yr (0.76 K for $\tau = 20$ yr). The improved agreement of the curves with the experimental data should be noted. It is felt that a time constant of about 30 years is reasonable for the equilibration by heat transfer between the pole area and the hemisphere, e.g. because of the slow response of ocean currents [28]. We mention that the warming of the SH is reported to be smaller than that for the northern hemisphere during the past three decades [25]. Our simulation nicely reproduces the difference of ~0.3 K in 2015 between the two hemispheres (for Northern Hemisphere see [3]). The accordance gives some support to the validity of the theoretical model.

Several authors pointed towards a pause of the global temperature rise in 2000 to 2010, presenting also potential reasons for this feature [29]. Our results are consistent with such a (small) effect buried somehow in the scatter of the experimental values. The Antarctic ice retreat in 1955 - 1979 dominates the minor ice growth since 1979 because of the slow equilibration of temperature changes in the SH, even beyond 2015. If the Antarctic sea ice area will not decrease in the coming years, the warming is likely to slow-down in the years 2015 to 2025. The phenomenon is illustrated by the inset in Fig. 2. The calculated warming for $\tau = 30$ yr (blue curve) is shown in the period 1975 to 2015 and extrapolated until 2030 (broken blue line). For the sea ice we assume to develop along the linear fit of Fig. 1, while for the GHG concentrations a growth with continuing slopes as in previous years is considered. Compared with the interval 2005 to 2015 the warming is predicted to decrease by approximately 33 % in 2025 (from 0.078 K in 2005-2015 to 0.058 K in 2015-2025). The major reason for the slowing-down is the long equilibration time for the ice retreat in 1955 to 1979. This feature dominates the opposite effect of the growing GHG. A possible effect of sun spot changes is omitted in the extrapolation. In order to reduce the experimental scatter of the temperature data (open grey circles in Fig. 2) running 3-year averages of the annual experimental values are plotted in the inset (see open blue circles). It is interesting to note the obviously systematic deviations of the experimental points from the computed curve in the inset.

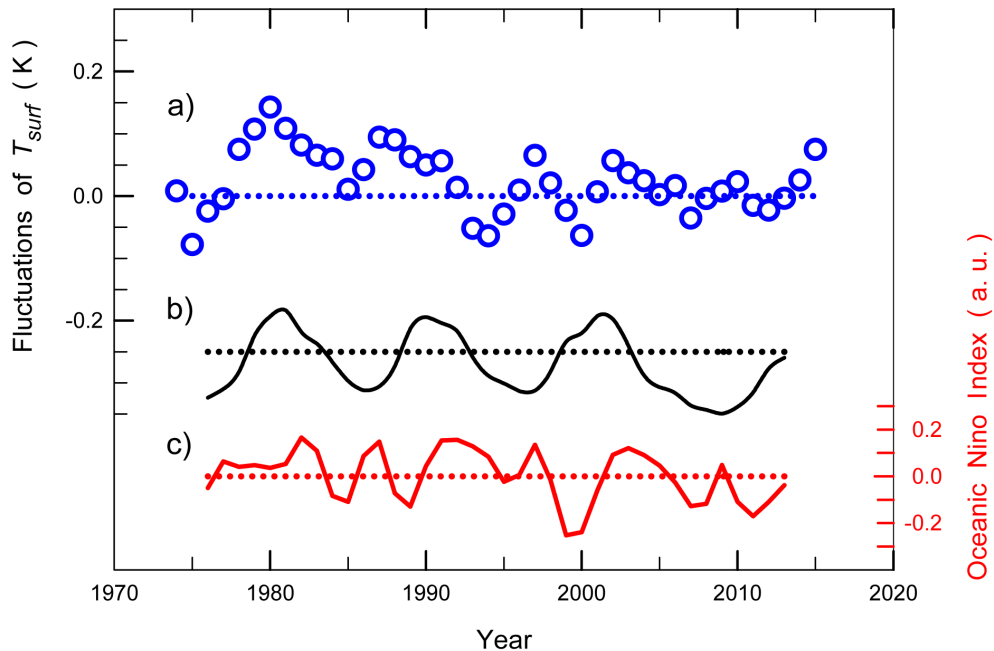

**Figure 3**: Fluctuations of the reported SH surface temperature, sun spot intensity and El Nino amplitude in 1975 - 2013. a) difference between the annual 3-year averages of $\Delta T_{surf,exp}$ and the calculated temperature rise $\Delta T_{surf}$ (blue open circles, left hand ordinate scale; compare inset of Fig. 2); b) surface temperature variation calculated from the reported oscillation of solar intensity [31,32] without a fitting parameter (black solid line, left hand ordinate scale); the data are down-shifted by -0.25 K for better visibility. c) Time evolution of the 3-year mean values of the reported Oceanic Nino Index (red solid line, r.h.s. ordinate scale) [33]. The similarity of some maxima and minima positions in a) to c) is noteworthy.

The finding is shown in more detail in Fig. 3. The difference between the blue experimental points (3-year averages) and the blue computed curve in the inset of Fig. 2 is plotted (open blue circles). We note several maxima and minima of the annual variations in the period 1975 to 2013. The question arises if a correlation with the fluctuations of the solar intensity (sunspot activity, [30]) and/or pacific temperature changes (El Nino observations) exists. Information of this kind is illustrated by Fig 3.

The straight dotted blue line in Fig. 3a refers to the blue curve in the inset of Fig. 2. It is interesting to see several clear maxima and minima (around +0.1 K and -0.05 K, respectively) [31,32]. Various authors pointed towards a correlation between sun spot activity and surface temperature [30]. In fact, a variation of the input solar intensity generates corresponding changes of $\Delta T_{surf}$ in our theoretical model. The results are plotted in the Fig. 3 b (solid black curve, left hand ordinate scale; note down-shift by -0.25 K for better visibility). The mean value of the solar intensity is represented by the dotted black line. There is no fitting parameter in the computation to derive the temperature maxima and minima of ±0.07 K that originate from fluctuations of the solar constant of ±0.5 W/m$^2$. Keeping in mind that there is still experimental error in the data, the present authors believe that the partial agreement between experimental points and the sun spot effect of Fig. 3b, e.g. the maxima around 1980 and 2000, gives some additional support to the validity of our 1D-model for surface temperature changes.

For comparison El Nino data [33] are also included in Fig. 3. It was reported that the striking peak of the observed surface temperature around 1942 is accompanied by an El Nino maximum [34]. driving strong pacific teleconnections. In Fig. 3c, three-year averages of the annual Oceanic Nino

Index [33] are presented for comparison (red curve, r.h.s. ordinate scale). Quite interestingly, some extremal positions agree to those in Fig. a). The comparison suggests a certain correlation between the El Nino amplitudes and the SH surface temperature. Of course, the feature does not provide a complete explanation for the SH temperature fluctuations over various decades.

In conclusion two dominant factors for the global warming of the southern hemisphere are discussed in the present paper: the retreat of Antarctic sea ice area and the growth of greenhouse gases since 1955. Using a simple 1-dimensional model, we estimate the contribution of ice-albedo feedback to the warming of the SH to be $0.5 \pm 0.1$ K in the years 1955 to 2015. For the simultaneous growth of the greenhouse gases ($H_2O$, $CO_2$, $CH_4$, $N_2O$) we calculate a temperature increase of $0.2 \pm 0.05$ K, derived directly from the spectroscopic properties of the GHG. A possible indirect effect is not included in this number, i.e. enhancement of the ice retreat by the greenhouse effect. The computed SH warming is in nice accordance with the reported mean temperature rise of $0.75 \pm 0.1$ K in 2015 relative to 1955. Our data also suggest interesting correlations between the annual fluctuations of solar intensity and El Nino observations on the one hand and the annual variability of the SH surface temperature on the other hand. Extrapolation of the sea ice and GHG data for the next decade suggests that the warming will slow down by 33 % in 2015-2025 as compared to the preceding decade.

APPENDIX 1: Estimate of the albedo change by the variation of Antarctic sea ice area

The data of Fig. 1 indicate a loss of average sea ice area of $3.95 \times 10^{12}$ m$^2$ for the years 1955 to 1979. For the subsequent period 1979 - 2015 the satellite data show that the maximum (minimum) sea ice area increased by $0.73 \times 10^{12}$ m$^2$ ($0.41 \times 10^{12}$ m$^2$) with a phase shift of ≈ 96 days for the seasonal rhythm relative to the sun (fitted numbers, see Fig. 1). As a result, the loss of average sea ice in 1955 to 2015 amounts to $3.38 \times 10^{12}$ m². Melting replaces snow-covered sea ice by a water surface lowering the backscattering by a factor of ≈ 0.5 [2,35]. Averaging over the daily and seasonal changes yields a further reduction factor of 0.164 for the mean solar input. Finally including the average transmission of 0.49 of the atmosphere in the melting region we estimate that the back-reflection of the SH surface into the universe decreases by $\Delta P_{relf} = -1.85 \times 10^{14}$ W in 2015 relative to 1955. As compared to the total solar input of the hemisphere of $P_{tot} = 8.73 \times 10^{16}$ W, we arrive at a relative change of $G_{rel} = \frac{\Delta P_{refl}}{P_{tot}} = -0.00212$. As compared to the global albedo of 0.300 this corresponds a relative decrease of ≈ -0.707 % in 2015. For the years 1955 to 2015 $G_{rel}$ is readily obtained as a function of time according to the annual ice data of Fig. 1. For delayed response of the total hemisphere to the changes in the polar region we introduce the mean albedo change Δ*albedo(t)* that is governed by the simple relaxation ansatz of Eq. 1:

$$\frac{d}{dt}\Delta albedo + \frac{\Delta albedo}{\tau} = \frac{G_{rel}(t)}{\tau} \qquad (1)$$

Here $\tau$ denotes the equilibration time. Eq. 1 is readily solved for Δ*albedo*(*t* = 1955) = 0 and decreases to -0.00181 in 2015 for $\tau$ = 30 yr (-0.00205 for $\tau$ = 20 yr). Our theoretical model responds quite linearly to small albedo changes with a scaling factor $\Delta T_{surf}/\Delta albedo \approx 247$ K. We mention that the albedo declines by loss of sea ice discussed here is lacking direct experimental verification. Changes of the global albedo of opposite signs (±0.02) were reported in the years 1985 – 1997 and 1997–2004 [36,37], or that the global albedo remained fairly constant during the past decades [38]. We propose that the pronounced rise of the reported surface temperature in Fig. 2 in 1970–2015 gives some experimental support for our estimate of the albedo decrease for the southern hemisphere.

APPENDIX 2: Theoretical model

In a 1-dimensional approach the change $\Delta T_{surf}$ of the mean temperature of the SH surface is calculated as a function of the concentrations $x_j$ of the GHG ($j$ = $CO_2$, $CH_4$ and $N_2O$) including water changes in the atmosphere, and the albedo. The model is based on the spectroscopic properties of the gases and empirical results, e.g. the albedo factor (*albedo* = 0.300 in 1880 - 1955) and the solar radiation input $S$ = 343.3 W/m². The atmosphere is represented by four layers with mean temperatures $T_j$ (j =1 – 4) that are self-consistently determined from thermal quasi-equilibrium between the solar input and the thermal emissions of the surface and atmospheric layers; i.e. energy conservation per unit time is assumed for input and output radiation into the universe. In the following subscript $j$ = 0 refers to the surface.

The surface is represented by a black radiator with emission intensity $\sigma \cdot T_0^4$ (Stefan-Boltzmann constant $\sigma$ [39]) while the atmospheric parts are treated as selective thermal emitters (subscript 0 refers to the surface). The spectral efficiencies $\eta_{ij}$ ($i$ = 0 - 4, $j$ = 1 - 4) of the atmospheric layers $j$ are calculated from the measured absorption properties of the greenhouse gases (including water) for the respective concentrations and temperatures $T_j$ without a fitting parameter. The temperature $T_i$ of the emitting surface ($i$ = 0 - 4) also comes in, since the spectral intensity distribution of the emission of the latter is temperature-dependent (compare Planck's formula) [39]. The net solar input to the surface is $A_0 = S \cdot (1 - albedo) - A_{atm}$ while $A_{atm}$ denotes the input loss in the atmosphere. Layer $j$ ($j$ = 1 - 4) obtains the fraction $A_j$, and $\sum_{j=1}^{4} A_j = A_{atm}$. The heat transport between the layers by non-radiative processes is included by parameters $B_j$ but is found to have little quantitative effect on the surface warming, since $A_{atm}$ is adjusted correspondingly to maintain $T_{surf} = 288.0$ K for year 1880 [22]. We arrive at five equations for $j = 0 - 4$:

$$\sum_{i=0}^{4} \eta_{ij} \cdot T_i^4 = f_j \cdot \eta_{jj} \cdot T_j^4 - \frac{A_j + B_j - B_{j+1}}{\sigma}. \qquad (2)$$

For a ready display of equations (2) we introduce formal parameters $\eta_{00} = 1, f_0 = 2, f_j = 3$ ($j > 0$), $B_0 = 0$, and $B_5 = 0$.

The molecular number densities of the GHG in the layers are calculated from a generalized barometric law. The water content in the atmosphere (integrated molecular number density) is taken to be $9.2 \times 10^{27}$ m$^{-2}$ (in 1880) for saturated vapor density. The amount of water follows the layer temperatures depending on the GHG concentrations of the individual year. An enhancement of $\Delta T_{surf}$ results by a factor of approximately 1.3 compared to constant water densities in the atmosphere. This number is somewhat smaller than reported by Wang et al. [40].

The GHG concentrations are taken from Refs. [24] and [25]. In 1880 we have $x_{CO2} = 290.4$ ppmv, $x_{CH4} = 1.6$ ppmv and $x_{N2O} = 0.3$ ppmv. Assuming equal absorption of the four atmospheric parts for $CO_2$, the thickness of the layers is estimated from a barometric formula to be 1.4, 1.9, and 2.8 km (layer 1 to 3, respectively), while the top layer (4) extends above 6.1 km into the universe. The numerical solution of equations shown above delivers the mean temperatures of respectively 285.2 K, 278.1 K, 268.3 K and 240.1 K, for layers 1 to 4. We have shown recently that the specific parameter values of $A_j$ and $B_j$ vary the layer temperatures of the atmosphere to some extent, but have little influence on the concentration dependence of the surface temperature [22]. Considering an analogous model with only three atmospheric layers we arrive at similar values for the surface temperature. We have also treated a 2D-approach by including the variation of the surface input with geographical latitude, obtaining similar results with changes of only a few percent (data not shown). It is concluded that our 1-dimensional 5-layer-model provides reliable data on the warming of the terrestrial surface.

AUTHOR INFORMATION

**Corresponding Authors**

*Email: hristo.iglev@ph.tum.de, alfred.laubereau@ph.tum.de,

**Notes**

The authors declare no competing financial interests.

REFERENCES

1. Bronwyn Wake; Polar response asymmetry. *Nat. Clim. Change* 4 (2014) 533.

2. Perovich D. K., Nghiem S. V., Markus T. and Schweiger A.; Seasonal evolution and interannual variability of the local solar energy absorbed by the Arctic sea ice-ocean system. *J. Geophys. Res.* 112 (2007) C03005.

3. Laubereau A. and Iglev H.; Arctic sea ice and the mean temperature of the northern hemisphere (2017); arXiv:1706.05835 [physics.geo-ph].

4. Flanner M. G., Shell K. M., Barlage M., Perovich D. K. and Tschudi M. A.; Radiative forcing and albedo feedback from northern hemisphere cryosphere between 1979 and 2008. *Nat. Geosci.* 4 (2011) 151.

5. Ding Q. et al.; Influence of high-latitude atmospheric circulation changes on summer-time Arctic sea ice. *Nat. Clim. Change* 7 (2017) 289.

6. Graversen R. G., and Wang M.; Polar amplification in a coupled climate model with locked albedo. *Clim. Dyn.* 33 (2009) 629.

7. National Academies of Sciences, Engineering, and Medicine; Antarctic sea ice variability in the Southern Ocean-climate system: proceedings of a workshop. (2017) DC: *The National Academies Press.* https://DOI.org/10.17226/24696.

8. Johanessen O. M. et al.; Satellites evidence for an Arctic sea ice cover in transformation. *Science* 286 (1999) 1937.

9. Turner J. and Comiso J.; Solve Antarctica's sea-ice puzzle. *Nature* 547 (2017) 275.

10. Cavalieri D. J. and Parkinson C. L.; 30-year satellite record reveals contrasting Arctic and Antarctic decadal sea ice variability. *Geophys. Res. Lett.* 30 (2003) 1970.

11. Turner J., Hosking J. S., Marshall G. J., Phillips T., and Bracegirdle T. J.; Antarctic sea ice increase consistent with intrinsic variability of the Amundsen Sea Low. *Clim. Dyn.* 46 (2016) 2391.

12. Comiso J. C., Kwok R., Seelye M., and Gordon A. L.; Variability and trends in sea ice extent and ice production in Ross Sea. *J. Geophys. Res.* 116 (2011) C04021.

13. Parkinson C. L. and Cavalieri D. J.; Antarctic sea ice variability and trends, 1979-2010. *The Cryosphere* 6 (2012) 871.

14. Holland P. R. and Kwok R.; Wind-driven trends in Antarctic sea-ice drift. Nature Geoscience 5 (2012) 872.

15. Screen J. A. and Simmons I.; The central role of diminishing sea ice in recent Arctic temperature amplification. *Nature* 464 (2010) 1334.

16. Cvijanovic I. and Caldeira K.; Atmospheric impacts on sea ice decline in $CO_2$ induced global warming. *Clim. Dyn.* 44 (2015) 1173.

17. Cavalieri D. J., Parkinson C. L.; 30-year satellite record reveals contrasting Arctic and Antarctic decadal sea ice variability. *Geophys. Res. Lett.* 30 (2003) 1970.

18. National Snow and Ice Data Center; All about sea ice (2016). https://nsidc.org/ cryosphere /seaice/characteristics/difference.html.

19. Rayner N. A. et al.; Global analyses of sea surface temperature, sea ice, and night marine air temperature since the late nineteenth century. *J. Geophys. Res.* 108 (2003) 4407.

20. De la Mare W. K.; Abrupt mid-twentieth-century decline in Antarctic sea ice extent from whaling records. *Nature* 389 (1997) 57.

21. Raith W. et al.; Bergmann-Schaefer Lehrbuch der Experimentalphysik: Erde und Planeten (Walter de Gruyter, Berlin 1997).

22. Onorato P., Mascheretti P., and Deambrosis, A.; "Home made" model to study the greenhouse effect and global warming. *Eur. J. Phys.* 32 (2011) 363.

23. Laubereau A., and Iglev H.; On the direct impact of the $CO_2$ concentration rise to the global warming. *Eur. Phys. Lett.* 104 (2013) 29001.

24. Blasing T. M.; Recent greenhouse gas concentrations (2016). cdiac.ornl.gov/pns/current-ghg.html.

25. Dlugokencky E., and Tans P.; Trends in atmospheric carbon dioxide. NOAA 2015, www.esrl.noaa/gmd/ccgg//trends/global.html.

26. NASA 2016; Goddard Institute for Space Studies: GISS Surface Temperature Analysis, http://data.giss.nasa.gov/gistemp/.

27. Myhre G. et al.; Anthropogenic and natural radiative forcing. In Climate Change (2013). The Physical Science Basis, eds Stocker T. F. et al. (Cambridge Univ. Press, 2013) p. 661

28. Goosse H. and Renssen H.; A simulated reduction in Antarctic sea-ice area since 1750: implications of the long memory of the ocean. *Int. J. Climatol.* 25 (2005) 569.

29. Guemas V., Doblas-Reyes F. J., Andreu-Burillo I., and Asif M.; Retrospective prediction of the global warming slowdown in the past decade. *Nat. Clim. Change* 3 (2013) 649.

30. Friis-Christensen E., and Lassen K.; Length of the solar cycle: an indicator of solar activity closely associated with climate. *Science* 254 (1991) 698.

31. Hansen J. et al.; Assessing "dangerous climate change": required reduction of carbon emissions to protect young people, future generations and nature. https://pubs.giss.nasa.gov. /docs/2013/ 2013_Hansen_ha085104.pdf.

32. Yeo K. L., Krivova N. A., Solanski S. K., and Glassmeier K. H.; Reconstruction of total and spectral irradiance from 1974 to 2013 based on KPVT, SOHO/MDI, and SDO/HMI observations. *Astronomy & Astrophysics* 570 (2014) 18.

33. Meehl G. A., Hu A., Santer B. D., and Xie S.-P.; Contribution of the interdecadal pacific oscillation to twentieth-century global surface temperature trends. *Nat. Clim. Change* 6 (2016) 1005.

34. Schneider D. P. and Steig E. J.; Ice cores record significant 1940s Antarctic warmth related to the climate variability. *Proc. Natl. Acad. Sci. USA* 105 (2008) 12154.

35. Kappas M.; Klimatologie (2009): Spektrum Akademischer Verlag, Heidelberg, p. 81.

36. Palle E. et al.; Earthshine and the earth's albedo: 2. Observations and simulations over 3 years. *J. Geophys. Res.* 108 (2003) 4710.

37. Palle E., Goode P. R., Motanes-Rodriguez P., and Koonin S. E.; Changes in earth's reflectance over the past two decades. *Science* 304 (2004) 1299.

38. Qiu J. et al.; Earthshine and the earth's albedo: 1. Earthshine observations and measurements of the lunar phase function for accurate measurements of the earth's bond albedo. *J. Geophys. Res.* 108 (2003) 4709.

39. Jenkins F. A. and White H. E.; Fundamentals of Optics (McGraw-Hill International Book Company) 1976.

40. Wang W. C., Yung Y. L., Lacis A. A., Mo T., and Hansen J. E.; Greenhouse effects due to man-made perturbations of trace gases. *Science* 195 (1976) 685.